\documentclass[useAMS,usenatbib]{mn2e}

\pdfoutput=1
\usepackage[english]{babel}
\usepackage[babel]{csquotes}
\usepackage{tikz}
\usepackage{etoolbox}
\usepackage{fancyhdr}
\usepackage{graphicx}
\usepackage{booktabs}
\usepackage{amsmath}
\usepackage{tensor}
\usepackage{placeins}
\usepackage{natbibmnfix}
\usepackage{epsfig}
\usepackage{aas_macros}
\usepackage{xcolor}
\usepackage{url}
\usepackage{amssymb}
\def\be{\begin{equation}}
\def\ee{\end{equation}}
\newcommand\code[1]{\textsc{\MakeLowercase{#1}}}

\newcommand\quotes[1]{``{#1}"}
\def\gsim{\lower.5ex\hbox{\gtsima}} 
\def\lsim{\lower.5ex\hbox{\ltsima}} 
\def\gtsima{$\; \buildrel > \over \sim \;$} 
\def\ltsima{$\; \buildrel < \over \sim \;$} 
\def\gsim{\lower.5ex\hbox{\gtsima}} 
\def\lsim{\lower.5ex\hbox{\ltsima}} 
\def\simgt{\lower.5ex\hbox{\gtsima}} 
\def\simlt{\lower.5ex\hbox{\ltsima}} 

\def\Msun{M_\odot}

\def\kms{{\rm km\,s}^{-1}\,}
\def\ergs{{\rm erg\,s}^{-1}\,}

\def\CII{\hbox{[C$\scriptstyle\rm II$]~}}
\def\NV{\hbox{N$\scriptstyle\rm V$}}
\def\HeII{\hbox{He$\scriptstyle\rm II$}}

\def\Mbh{M_\bullet}

\title[Black holes in Lyman Break Galaxies]{Massive black holes in high-redshift Lyman Break Galaxies}
\author[M.C. Orofino, A. Ferrara, S. Gallerani] {M.C. Orofino, A. Ferrara, S. Gallerani
\\
\noindent Scuola Normale Superiore, Piazza dei Cavalieri 7, I-56126 Pisa, Italy\\}

\date{}
\def\LaTeX{L\kern-.36em\raise.3ex\hbox{a}\kern-.15em
    T\kern-.1667em\lower.7ex\hbox{E}\kern-.125emX}

\begin{document}
\maketitle

\begin{abstract}
Several evidences indicate that Lyman Break Galaxies (LBG) in the Epoch of Reionization (redshift $z>6$) might host massive black holes (MBH). We address this question by using a merger-tree model combined with tight constraints from the 7 Ms Chandra survey, and the known high-$z$ super-MBH population. We find that a typical LBG with $M_{\rm UV}=-22$ residing in a $M_h\approx 10^{12} M_\odot$ halo at $z=6$ host a MBH with mass $M_\bullet \approx 2\times 10^8 M_\odot$. Depending on the fraction, $f_{\rm seed}$, of early halos planted with a direct collapse black hole seed ($M_{\rm seed}=10^5 M_\odot$), the model suggests two possible scenarios: (a) if $f_{\rm seed}=1$, MBH in LBGs mostly grow by merging, and must accrete at a low ($\lambda_E\simeq 10^{-3}$) Eddington ratio not to exceed the experimental X-ray luminosity upper bound $L_X^* = 10^{42.5} {\rm erg\, s}^{-1}$; (b) if $f_{\rm seed}=0.05$ accretion dominates ($\lambda_E\simeq 0.22$), and MBH emission in LBGs must be heavily obscured. In both scenarios the UV luminosity function is largely dominated by stellar emission up to very bright mag, $M_{\rm UV} \simgt -23$, with BH emission playing a subdominant role. Scenario (a) poses extremely challenging, and possibly unphysical, requirements on DCBH formation. Scenario (b) entails testable implications on the physical properties of LBGs involving the FIR luminosity, emission lines, and presence of outflows.
\end{abstract}


\section{Introduction}
The presence of massive black holes (MBH, $\Mbh \simeq 10^{7-8} \Msun$) in typical Lyman Break Galaxies (LBGs) in the Epoch of Reionization (EoR) has become a very pressing question in early galaxy and black hole (co-)evolution. These objects might hold the key to understand at least three fundamental issues: (a) do MBHs in LBGs represent the progenitor population of supermassive black holes (SMBH, $\Mbh \simgt 10^{9} \Msun$) powering the brightest quasars (QSO)? (b) what can we learn about black hole seeds and growth of these compact objects? (c) do they affect, and by what physical mechanisms, the properties and evolution of early galaxies and even large scale structure, e.g., contributing to the reionization and metal enrichment of the intergalactic medium?  

At present, we have collected significant statistics and luminosity functions of a large sample of high-$z$ galaxies over a wide span of magnitudes, thanks to space-born surveys, such as (i) the Hubble Ultra Deep
Field, and the eXtreme Deep Field, exploiting the power of WFC3 onboard the Hubble Space Telescope
\citep{2010ApJ...725L.150O, 2011ApJ...737...90B, bouwens2015, mcl2015yCat..74322696M, 2017bouwApJ...843..129B};  
(ii) the Hubble CLASH lensing surveys
\citep{atek2015ApJ...800...18A, mcleod2016MNRAS.459.3812M,  liver2017yCat..18350113L}; (iii) X-ray surveys \citep{suh2019, calhau2020}. These endevours are complemented by a number of ground-based surveys \citep{bradley2012ApJ...760..108B, bowl2015, bouwens2016ApJ...830...67B,fogasy2020}.
Although the detected galaxies are seen within $\simlt 1$ Gyr from the Big Bang, some of them have already built-up large stellar masses and appear as evolved systems, containing an almost solar abundance of heavy elements and dust. For recent reviews on these topics we defer the reader to \citet{dayal2018, Maiolino19}.

{
In lower-$z$ galaxies there is evidence for a relation between stellar mass and the massive black holes harbored at their centers \citep{Kormendy13, Heckman14, volonteri16}. This connection is not yet fully established at high-$z$. Some cosmological hydrodynamical simulations\footnote{
For a comparative study on large scale cosmological hydrodynamical simulations we refer to
\cite{hab2020arXiv200610094H}. Such work focuses on the mass properties of SMBH and on their relation
with the stellar mass of the host galaxies in six different simulations. It is worth noting that while all the simulations generate a $M_\bullet - M_*$ relation in
general accordance with observations, some simulations are in tension with the data for low-mass BHs $M_\bullet < 10^{7.5} M_\odot$.    } such as Horizon-AGN \citep{volonteri:horizon} and BlueTides \citep{huang:bt} show no significant evolution in the relation up to $z\sim 8$. { The same conclusion is found 
by \cite{marshall20}} with a semi-analytical model that highlights minimal evolution in the black hole–bulge and black hole–total stellar mass relations out to $z=8$ \cite[see also][]{Lupi2019}.
Other results from hydrodynamical simulations \citep{Khandai2012, Barai18} and semi-analytical models \citep[e.g.][]{Lamastra2010} of $z\sim 6$ SMBHs show deviations 
from the local relation \citep{Kormendy13}. 
These results seem to be confirmed by
a handful of high$-z$ observations \citep{Wang10, Targett12, Willott15, Pensabene20} available to date.
They show an over-massive black hole trend with respect to the host stellar mass. 
In particular, the analysis by \cite{Targett12} at $z\sim 4$ shows a fast growth of the BH-to-stellar mass ratio with redshift as $\propto (1+z)^{1.4-2.0}$. 
However, these works are still debated and not conclusive.
Specifically, \cite{salvia2007ApJ...662..131S} concluded that apparent evolution of the local law can be due to observational biases.

Observations} are jeopardized by several difficulties. The standard direct { X-ray detection of AGN} technique, widely applied at lower redshifts, becomes very challenging for these remote and intrinsically faint ({ or obscured, \citealt{tre2019MNRAS.487..819T, ni2020MNRAS.495.2135N}}) objects. Fortunately, new results pushing instrumental capabilities to their very limits have been nevertheless obtained: using the 7 Ms Chandra survey \citep{Vito18, Cowie20} have derived very stringent and useful constraints on the early MBH population. In this paper we will extensively made use of these constraints to calibrate and anchor our models.
As the UV emission from MBH is likely swamped by stellar light emitted by the galaxy, observations in this band can be hardly conclusive about the presence of a central black hole in LBGs. Faint Active Galactic Nuclei (AGN) might also be present within Lyman Alpha Emitters (LAE) as recently shown by \citet{calhau2020} (see also \citet{Haro20}). These authors studied the X-ray and radio properties of about 4000 LAEs at $2.2 < z < 6$ from the SC4K survey in the COSMOS field. They detect 6.8\% (3.2\%) of these sources in the X-rays (radio). The interpretation of these results relies on the existence of a population of extremely faint/obscure AGN that escape even the deepest X-ray searches, but are potentially detectable in radio emission.  

In spite of these difficulties, MBH might be caught by searching for the unique features they imprint on the observed properties of the host galaxy.  These indirect probes might then allow us to reliably answer the questions outlined at the beginning.  Among other possibilities, there are at least three indirect but clear smoking guns of the presence of a hidden MBH in a LBG. 

The first is the infrared emission from an accreting MBH. MBH in LBG are either quiescent or heavily obscured by dust. In this second scenario strong IR emission is expected. Both the IR peak wavelength and intensity strongly depend on the spatial distribution of dust around the BH, and therefore on the galaxy/AGN morphology\footnote{ However, IR-observation of AGN in dwarf galaxies is particularly difficult as star formation in these systems is capable of heating dust in such a way that mimics the infrared colors of more luminous AGNs \citep{hain2016ApJ...832..119H}.}. The second probe are UV emission lines. UV emission lines such as \HeII\, 1640~\AA~and \NV\, 1240~\AA~are good tracers of a hard radiation field { \citep{Pallottini15, stark2015MNRAS.454.1393S, feltre2016MNRAS.456.3354F, volo2017ApJ...849..155V, Laporte17}} and have been systematically used to study AGNs \citep{Dietrich02}. Finally, MBH are in principle capable to launch powerful outflows { \citep{Costa2014,Costa2015,Barai18, ni2020MNRAS.495.2135N}}. In a dust-obscured AGN, in fact, dust opacity boosts radiation pressure efficiency well above the level expected from pure electron scattering \citep{Fabian08}. Outflows, in turn, might profoundly affect galaxy morphology, star formation, escape of ionizing photons and metal enrichment of the circumgalactic and intergalactic medium. Although at least some of these effects are degenerate with the star formation activity \citep{Dayal10, Pizzati20} or PopIII emission \citep{Pallottini15}, their combination can uniquely pinpoint the presence of a MBH. Furthermore, signatures of outflowing gas have been recently found in several $z>5$ galaxies \citep{Gallerani18, Sugahara19, Fujimoto19, Ginolfi20}. These outflows may be possibly powered by a yet undetected accreting MBH.

Here we use available data in combination with simple but robust semi-analytical models, similar to previous works by \cite{tanaka2009,petri2012, tanaka2014CQGra..31x4005T}, to study the mass and luminosity of MBH as a function of the host halo mass. With this approach we aim to assess whether LBGs at $z=6$ host MBH, determining the MBH mass and Eddington ratio and preliminarily appraise their impact on the galaxy properties. 

The paper is organized as follows\footnote{Throughout the paper, we assume a flat Universe with the following cosmological parameters:  $\Omega_{\rm M}h^2 = 0.1428$, $\Omega_{\Lambda} = 1- \Omega_{\rm M}$, and $\Omega_{\rm B}h^2 = 0.02233$,  $h=67.32$, $\sigma_8=0.8101$, where $\Omega_{\rm M}$, $\Omega_{\Lambda}$, $\Omega_{\rm B}$ are the total matter, vacuum, and baryonic densities, in units of the critical density; $h$ is the Hubble constant in units of $100\,\kms$, and $\sigma_8$ is the late-time fluctuation amplitude parameter \citep{Planck18}.}.  Sec. \ref{sec:meth} describes the merger tree, seeding and growth prescriptions, along with the observational constraints we impose. In Sec. \ref{sec:BH-halo} we derive the BH-halo mass relation, and in Sec. \ref{sec:lum} we use it to compute the combined galaxy-AGN luminosity function for two different scenarios. Sec. \ref{sec:impl} contains the implications for LBGs, including FIR luminosity, emission lines, and presence of outflows. Finally, a summary is given in Sec. \ref{sec:summary}.

\section{Method}\label{sec:meth}
We run merger trees by using the public code\footnote{\url{star-www.dur.ac.uk/~cole/merger_trees/}} described in \cite{mt}.
Our initial goal is the derivation of the relation between the mass of the BH and the host dark matter halo. We generate merger trees with root halos of mass in the range $M_h=10^{10.6-13} M_\odot $ at $z=6$. Then we seed the leafs with BHs, and follow their accretion and merging down to the root. Our method is similar to the one used by \cite{tanaka2009}, and is best suited to derive the most probable BH mass hosted by a halo of known mass, along with its expected variance. 

We specialize our analysis to redshift $z=6$, where {
 observational data  on UV and X-ray luminosity of AGN are available and can provide indirect constraints on the black hole-galaxy relation.} Our aim is to study the BH final mass and accretion rate at that epoch, even for those BH masses that are smaller than observed. We then require that the combined BH-galaxy luminosity functions satisfy the available constraints from deep UV/X-ray surveys, and use the results to make predictions for future early galaxies observations. 

\subsection{Merger Tree setup and parameters}
\cite{mt} algorithm follows the formation history of dark matter halos using the extended Press-Schechter theory. It has been shown to be in accurate agreement with the conditional mass functions found from $\Lambda$CDM Millennium N-body simulations. For further details on the code we refer the reader to \cite{mt}. 

We sample $40$ different final halo masses equally spaced in log space from $M_{\rm min} = 10^{10.6} M_\odot$ to $M_{\rm max} =10^{13} M_\odot $ and run $\simgt 100$ merger trees for each mass to achieve a significant statistics. The mass resolution of the merger trees is $M_{\rm res} = 5\times 10^7 M_\odot$; they follow the cosmic evolution from $z=30$ to $z=6$.
For each of the trees we computed the BH mass inside the halos, following accretion and merging history from the high-$z$ leafs to the $z = 6$ root. The seeding and accretion prescriptions are described in the following.

\subsubsection{Seeding}\label{seed}
Seeding prescriptions employing stellar mass BH ($\Mbh \simlt 10^2 M_\odot$) seeds are known to face difficulties in explaining the observed $10^9M_\odot$ SMBHs at $z=6$ \citep{alvarez:2009, paper1}. Such models have to resort to prolonged super-Eddington accretion phases \citep{madau:2014, volonteri:2015, aversa2015ApJ...810...74A, lupi:2016, regan2019MNRAS.486.3892R, takeo2020arXiv200207187T}, in contrast with { models in which radiation pressure regulates gas infall}, such as e.g. \citet{park2012PhDT.......155P, park_ricotti, toyo2020arXiv200208017T, sugi2020arXiv200305625S}.

We use instead a seeding prescription based on intermediate mass $\Mbh \approx 10^5 M_\odot$ black holes (IMBH). These seeds represent the possible outcome of two direct formation scenarios: (i) monolithic collapse of the gas in $\rm H_2$-free primordial halos (direct collapse BH, see, e.g. \citealt{rees:1984};  { \citealt{bro2003ApJ...596...34B};}
\citealt{maye2010Natur.466.1082M, ferrara2014, mayer2015ApJ...810...51M})\footnote{ In particular, the BH formation scenario developed in \cite{maye2010Natur.466.1082M, mayer2015ApJ...810...51M} is known as \textit{cold direct collapse} and relies on merging galaxy cores.}, and (ii) heavy seeds formation in low-metallicity, dense stellar clusters where, due to energy equipartition, the most massive members tend to sink towards the center \citep{spitzer:1969,vishniac:1978, begelman:1978,lee:1987,quinlan:1990, omukai:2008, devecchi:2009, devecchi2010MNRAS.409.1057D, devecchi2012MNRAS.421.1465D, mehr2019ApJ...887..195M,  boco2020ApJ...891...94B}. { We refer to \citet{latif2016}, \citet{mez2017IJMPD..2630021M} and \citet{Ina20} for recent reviews on this topic.}

\cite{ferrara2014} studied the IMBH initial mass function and host halo properties. They concluded that a { good}  prescription is to seed halos of mass $7.5 < \log (M_h/ M_\odot) < 8$ in the redshift range $8 < z < 17$ with IMBH of mass $4.75 < \log (\Mbh/ M_\odot) < 6.25$. We then adopt this prescription, but for simplicity assume a single\footnote{ As a test, we checked that a random scatter in the seed mass range $(0.2-1.8) \times 10^5 M_\odot$ introduces variations $<5$\% in the results.} mass value $\Mbh = 10^5 M_\odot$.

We also note that the above results represent only a necessary condition for the formation of IMBH. The prescription implicitly assumes that the halo is illuminated by a sufficiently strong UV Lyman-Werner (LW; $11.2-13.6$ eV) intensity $J_{LW} > J_{LW}^*$ so to prevent molecular hydrogen formation during the collapse. The precise value of the intensity threshold, $J_{LW}^* \approx (30-1000)\times 10^{-21} {\rm erg\, s}^{-1} {\rm cm}^{-2} {\rm Hz}^{-1} {\rm sr}^{-1}$, depends on radiative transfer, chemistry and spectral shape of the sources, and it is only approximately known \citep{ferrara2014,sugi2014MNRAS.445..544S,aga2015MNRAS.446..160A, agarwal2016MNRAS.459.4209A}.
We condense this uncertainty in the $f_{\rm seed}$ parameter expressing the fraction of potential host candidate halos that actually meet the above illumination condition, and therefore are seeded with an IMBH. 

{ To summarize our seeding procedure: we planted a seed of $10^5 M_\odot$ in a fraction $f_{seed}$ of the merger-tree leaves that have mass $7.5 < \log (M_h/ M_\odot) < 8$ in the redshift range $8 < z < 17$; $f_{seed}$ is constant over such mass/redshift ranges.}
{ As we will see in Sec. \ref{sec:BH-halo}, we distinguish two different scenarios: in the first one, all the IMBH host halo candidates are planted with a seed; in the other, only a small { randomly selected} fraction of the candidates hosts a seed.
Different values of the fraction of seeded halos $f_{seed}$ lead to  qualitatively different BH build-up scenarios:
$f_{seed} \approx 1$ implies that the bulk of the BH mass is gained by seed-merging, while in the lower $f_{seed}$ case the BH mass is mainly gained by direct gas accretion. Note that $f_{seed}$ is a free parameter of our model.
}

\subsubsection{Growth}\label{growth}
Implanted BH seeds can grow via two distinct channels: BH-BH merger and direct { gas} accretion. We assume that every halo merger results in an instantaneous BH merging. This is justified by previous calculations \citep{Armitage05, Volonteri06, Colpi14} which showed that coalescence is very rapid due to the fact that both viscous dissipation in the surrounding accretion disk, and energy loss due to gravitational wave emission have timescales much shorter than the Hubble time ($\approx 1$ Gyr) at $z=6$.

For simplicity, we do not consider gravitational–radiation induced recoil \citep{Merritt09,Devecchi09} neither.  This assumption is partly justified by the findings of \citet{Schnittman07} indicating that the typical velocities for gravitational recoil (or “kick”)
are of the order of $100\,\kms$. These values are lower than the escape velocity from the typical halos we are interested in ($M_h \approx 10^{12} M_\odot$ corresponding to $v_e = 422\, \kms$).

However,
 { other works (e.g. \citealt{tambure2017MNRAS.464.2952T, pfister2019MNRAS.486..101P}; \citealt{bortolas}) deem these assumptions as too optimistic, 
as some BHs might be kicked off from the galaxies during the initial growth stages or because the merging time might not be negligibly short. This might lead to an overestimate of the merger efficiency, and hence of the final MBH mass { in  merging-dominated scenarios. However, this issue does not affect the results of accretion-dominated scenarios. }
}

In between two merger episodes, we allow BHs to grow by direct accretion at a fraction $\lambda_E$ of their Eddington rate, $\dot{M}_{\bullet} = \lambda_E \dot{M}_E$, where
\begin{equation}
         \dot{M}_E \equiv \frac{L_E}{\eta c^2} =  2.5 \times 10^{-8} \left(\frac{\Mbh}{M_\odot}\right) \left(\frac{0.1}{\eta}\right) M_\odot\rm yr^{-1},
\end{equation}
where $L_E = 1.5 \times 10^{38}(\Mbh/M_\odot)\, \rm erg\, s^{-1}$ is the Eddington luminosity. The allowed values for the matter-radiation conversion efficiency, $\eta$, range from 0.054 for non-rotating Schwarzschild BHs to 0.42 for maximally rotating Kerr BHs \citep{Shapiro83}. Following the arguments given in \citet{Marconi04} we will take $\eta=0.1$ in the following.
The Eddington-ratio $\lambda_E$ is a free parameter of the model, and it is further discussed in the next Sections;
{ in particular, we refer to Sec. \ref{ssec:lambdae}, where the calculation of the Eddington-ratio is summarized.}

\subsection{BH Luminosity}
From the assumptions made in the previous Section it follows that the bolometric luminosity of the BH is simply given by 
\begin{equation}
    L= \lambda_E(M_\bullet) L_E.
    \label{Lbol}
\end{equation}
In the previous equation we have highlighted the likely possibility that the Eddington ratio is a function of BH mass. This function is not yet specified in our model. In the next Section we will discuss the constraints on $\lambda_E$ descending from available observational data. We will then explore the effects of different $\lambda_E$ prescriptions.

For later use we will need to calculate the X-ray ($0.5-2$ keV band) and UV (at 1450~\AA) luminosity of the black holes. These can be directly obtained from the bolometric luminosity by applying the appropriate bolometric corrections, i.e.
\begin{equation}
    L_i = f_i(L)  L, 
    \label{Lband}
\end{equation}
with $L$ given by eq. \ref{Lbol}, and $i = {\rm X, UV}$. { For both $f_{\rm X}$ and $f_{\rm UV}$ we use the luminosity-dependent fit by \citet{shen2020} (see their Fig.2)}. For example, for $L=10^{46} {\rm erg\, s}^{-1}$ they find $f_X=0.02$ (soft band) and $f_{\rm UV}=0.2$, respectively.

\subsection{Observational constraints}\label{obscon}

In order to constrain { direct accretion efficiency and obscuration (see Sec. \ref{S1} and Sec. \ref{S2}) }, we use two types of experimental constraints available at $z\simeq 6$. The first one comes from the abundance and luminosity of SMBH. According to $\Lambda$CDM \citep{shethtormen, Mo02, Mo10} and for the adopted cosmology, the comoving density\footnote{The precise value of $n_{max}$ is somewhat sensitive to the cosmological parameters, particularly $\sigma_8$, and the transfer function used. As the primary goal of this work is to study MBH in LBGs, this uncertainty has virtually no impact on our results. To compute the halo mass function we have used the public code { \code{HMFcalc} \citep{mfpaper}}, available at \url{hmf.icrar.org/hmf\_finder/form/create/}.} of the most massive halos in our merger tree, $M_{max}=10^{13} M_\odot$, at $z=6$ is $n_{max}\simeq 10^{-9}\, \rm Mpc^{-3}$.  According to the results of the Sloan Digital Sky Survey \citep{Jiang09} this abundance corresponds to a UV magnitude (at 1450~\AA) in the range $-27<M_{\rm UV}<-25$. In this work we will use $M_{\rm UV} \simeq -26$  or $L_{\rm UV} \simeq 2\times 10^{46} {\rm erg\, s}^{-1}$  or $L \simeq 10^{47} {\rm erg\, s}^{-1}$. This sets a constrain on the product $\lambda_E M_\bullet = 6\times 10^8 M_\odot$. As measurements of the virial SMBH mass using MgII line width of individual SDSS sources and other high-$z$ quasars \citep{Willott07, Kurk07, mortlock:2011} indicate $M_\bullet \simeq (1-3)\times 10^9 M_\odot$, one can conclude that $\lambda_E \approx 0.2-0.6$ in the supermassive regime.

The second constraint comes from X-ray observations. \cite{Cowie20} searched for high-redshift ($z > 4.5$) X-ray AGN in the deep central region of the 7 Ms Chandra Deep Field-South (CDFs) X-ray image. They put a tight\footnote{We recall that the $L_X^*$ value is so low that it could be  produced purely by High Mass X-ray Binaries and hot ISM  in a galaxy forming stars at a rate of $\simeq 150 M_\odot {\rm yr}^{-1}$ \citep{Mineo14,Das17}; { see also Sec. \ref{ssec:gallum}.} } upper limit, $n_{X} = 10^{-5} \rm Mpc^{-3}$, on the comoving density of $z\simeq 6$ X-ray sources with $L_X > L_X^*=10^{42.5} \rm erg \, s^{-1}$ in the 0.5-2 keV band.

Paralleling the previous procedure, this abundance corresponds to  
a halo mass $M_h \simeq 10^{12} M_\odot$. Note that this halo mass scale is typical of $z=6$ LBGs with $M_{UV}\simeq -22$  (\cite{Behroozi19} and Fig. \ref{fig:lumfun}). CDFs observations then set a constrain $\lambda_E M_\bullet \simeq 10^6 M_\odot$ for BH hosted by LBGs at the end of the Epoch of Reionization. We should also allow for the possibility that a fraction of the emitted light by the AGN is absorbed by dust and gas in the vicinity of the black hole. When appropriate, we denote this fraction as $f_{\rm abs}$ in the appropriate X-ray or UV band. 

%
%
\begin{figure*}
\vspace{+0\baselineskip}
{
\includegraphics[width=1.\textwidth]{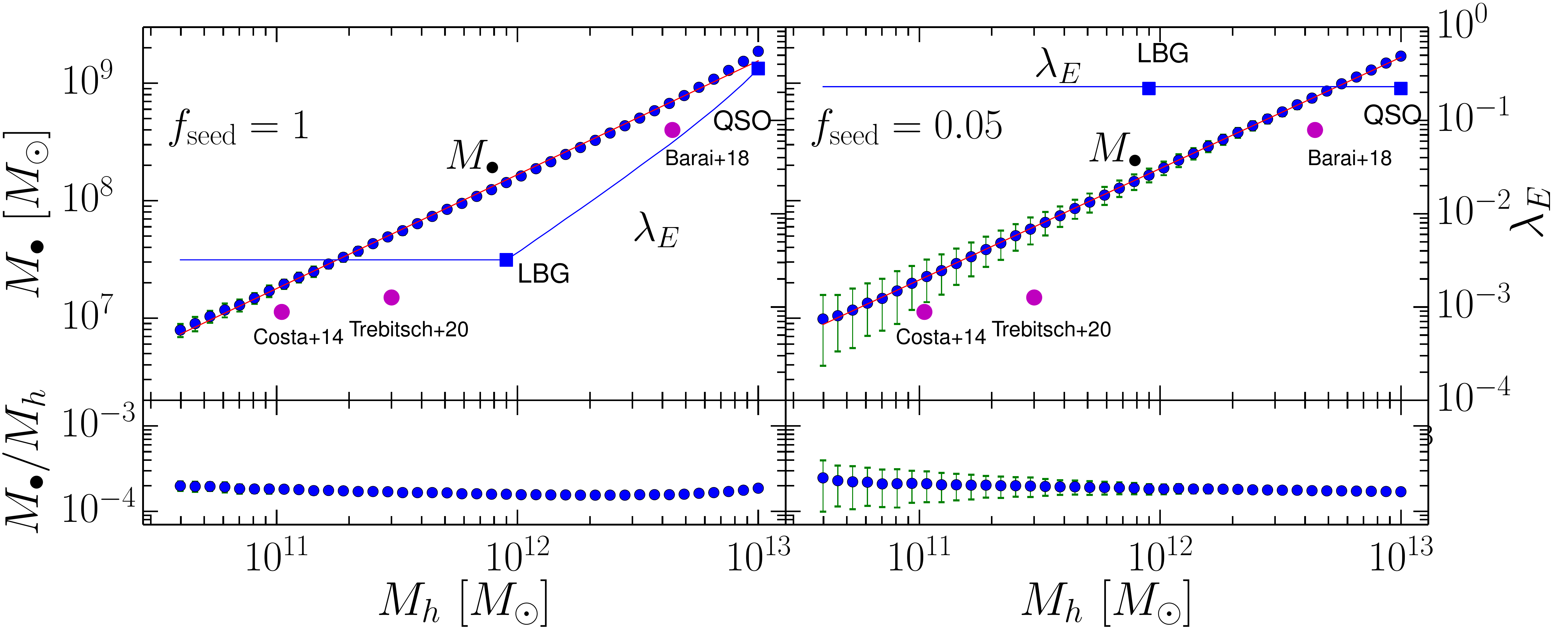}}

\caption{Halo vs massive black hole mass relation { at $z=6$}. \textit{Left panel}: Results of the merger tree simulations (blue points) for scenario S1 ($f_{\rm seed}=1$). Variance of individual points is evaluated from $\approx 100$ merger tree realizations performed per halo mass bin. The red line represents the best fit (see text for an analytical expression) to the data. The Eddington ratio, $\lambda_E$, required to match CDFs and QSO abundance data is shown by the blue line. Also shown is the result of numerical simulations { by \citet{costapunto2014MNRAS.439.2146C, Barai18,  Trebitsch20}: the predicted trend is consistent with these works, although our BH masses are slightly larger}. The bottom panel shows the $M_\bullet/M_h$ ratio across the halo mass range.
\textit{Right}: Same for S2 ($f_{\rm seed}=0.05$). 
}
\label{Fig01}
\end{figure*}

\section{Constrained Halo-BH relation}\label{sec:BH-halo}
As already mentioned, the seeding prescription we are using (see \ref{seed}) represents only a necessary condition for the formation of a IMBH by direct collapse. In fact, not all the candidate host halos might be illuminated by a sufficiently strong LW flux such to prevent H$_2$ formation, leading to detrimental cooling fragmentation during the collapse. Given this uncertainty, we explore in the following two different scenarios, S1 and S2.

\begin{itemize}
    \item\textit{Maximal seeding} [Scenario S1]. This scenario assumes that all the candidates IMBH host halos are exposed to a sufficiently high LW radiation field coming either from a nearby galaxy or the cosmological background. This implies a fraction $f_{\rm seed}=1$ of seeded halos in the leafs.
    \item\textit{Inefficient seeding} [Scenario S2] This scenario envisages a inhomogeneous LW background, as predicted by { most   studies (e.g. \citealt{Yue14})} in which only a small fraction of the putative host halos can form a IMBH. Guided by these findings, we seed a fraction $f_{\rm seed}=0.05$ of the leafs.
    { Due to the likely inhomogeneous topology of the LW background, S2 scenario seems largely favored.}
    
\end{itemize}
As we will see in the following, the main difference between the two scenarios is the relative importance of merging and accretion for the BH growth. In S1 the bulk of the final MBH mass is already made up by the seeds, and therefore limited accretion is required. In S2, instead,  the initial seed mass is decreased by 20 times, and therefore growth must occur largely by accretion. We discuss the implications of these two scenarios in the following.

{In this work we assume that BHs are active at all times, i.e. we do not introduce a duty cycle; however, we introduce obscuration in model S2 to satisfy the observational constraints. As obscuration and duty cycle are known to be degenerate \citep{shankar, Chen:2018, tremax2019}, the two possibilities can be disentangled only with the help from ancillary data such as IR observations or clustering experiments.}

{
\subsection{ Constraining the Eddington parameter}
\label{ssec:lambdae}
 In this Section we detail the procedure used to compute $\lambda_E$ in S1 and S2.

\begin{itemize}
    \item {\it Model S1}. In this model, growth is dominated by BH seed merging and direct accretion is significant only for SMBHs. We initialize the model with $\lambda_E=0$ and computed BH masses accordingly; $\lambda_E$ is then constrained for $10^{12} M_\odot$ and $10^{13} M_\odot$ halos with the observed X-ray luminosity. We find that, with S1 seeding assumptions, accretion is negligible for $10^{12} M_\odot$ halos, and it does not significantly affect the final BH mass. However, it becomes important for SMBH. To determine it, we run the model again iteratively with the new values of $\lambda_E$, { linearly interpolated in log-log scale} as we explain in Sec. \ref{S1}. At the end of each run, we used the final BH masses to compute the new values of $\lambda_E$ that comply with the $L_X$ constraints. We stopped the iteration when the change in LBG and SMBH masses was smaller than simulation variance.
    
    { Of course, the simplest prescription would be a constant $\lambda_E$, but this assumption is not suitable for model S1: indeed, in order to simultaneously comply with both the QSO and the LBG observational constraints, the Eddington-ratio should be relatively high and significant obscuration must be invoked.
We remind that in this scenario the bulk of the BH mass is gained by merging and that, without obscuration, the BH masses that we compute are $\approx 10^{-4} M_h$. Including obscuration and increasing $\lambda_E$ would lead to much higher final BH masses; most importantly, we would  overshoot SMBH masses. Finally, we point out that obscuration is degenerate with $\lambda_E$ and cannot be constrained independently.}

    \item {\it Model S2}. In model S2, $\lambda_E$ is constant. We fix $\lambda_E$ in order to produce in the most massive halo the same BH mass as in S1.
    {
    We stress that in this scenario $\lambda_E$ has to be relatively high to produce the supermassive black holes observed in quasars. This forces the introduction of obscuration in the model, thus justifying the simple prescription of a constant $\lambda_E$.}
    
\end{itemize}
}

{
Model S1 is an extreme and limiting case of BH accretion modeling as the final BH mass is dominated by seed mergers, 
and no obscuration has been introduced. In principle, it would be possible to modify the model in other reasonable ways, e.g. by introducing obscuration in LBG galaxies and adjust $\lambda_E$ accordingly. For simplicity, in this work, we     concentrate on the simplest scenarios in which BH growth is either accretion- or merger-dominated. However, most of the implications of our results will be investigated for model S2 which we consider as the fiducial one.
}

\subsection{Maximal seeding scenario}\label{S1}
In the S1 scenario $f_{\rm seed}=1$. Our model in this case predicts a tight, almost linear relation between the MBH and halo mass (Fig. \ref{Fig01}) { whose best fit is $\log \Mbh = -3.4^{+0.1}_{-0.1} + 0.97^{+0.01}_{-0.01}\log M_h$.}  Note that the variance in the relation, evaluated from the $\approx 100$ merger tree realizations performed per halo mass bin, is very small. 

Halos corresponding to typical LBGs at $z=6$ ($M_h \simeq 10^{12} M_\odot$) are predicted to host BHs with mass $M_\bullet = 2\times 10^8M_\odot$. Such ratio, $M_\bullet/M_h \approx 2 \times 10^{-4}$, is approximately constant in the entire BH mass range, as depicted in the lower panel of Fig. \ref{Fig01}; this value is significantly smaller than the seeding one, $0.001-0.03$. 

If we convert the halo mass into stellar mass using the models by \citet{Behroozi19} we obtain $M_\star=1.3\times 10^{10}\Msun$, corresponding to a BH/stellar mass ratio  ${\cal R}=0.015$. While higher than the local value, ${\cal R}=0.0037$ -- see eq. 10 of \citet{Kormendy13} -- yet this value is consistent with high-$z$ determinations \citep{Wang10, Targett12, Willott15}. Although at $z=6$ the observational relation is affected by a large scatter \citep{Pensabene20} { and observational bias}, our result confirms that BHs grew faster than their host stellar counterpart. These conclusions hold for both S1 and S2.

{ 
As $\lambda_E$ governs also the growth of the black hole ($d\log M_\bullet \propto \lambda_E$) we impose the SMBH abundance/mass and X-ray luminosity limits  according to  the  deep  central  region  of  the  7 Ms CDFs X-ray image (\citealt{Cowie20}, Sec. \ref{obscon}) into our merger tree by linearly interpolating { in log-log space} $\lambda_E$ in the halo mass range $10^{12-13} M_\odot$. For smaller halos we keep $\lambda_E \equiv {\rm const.}$ (see Fig. \ref{Fig01}); since this value is very low, $\simeq 10^{-3}$, implying a very inefficient accretion, our conclusions are weakly affected by a different assumption on the shape of $\lambda_E$ for halos $M_h \simlt 10^{12} M_\odot$. We then solve the problem by iteration.
The key challenge is to produce the SMBHs powering quasars, at the same time preventing MBHs to become too luminous.
}
{ $\lambda_E$ affects both $M_\bullet$ and the luminosity (in particular, the X-ray luminosity $L_X$).
The higher $\lambda_E$, the higher $L_X$, since $L_X$ is a monotonic function of $\lambda_E$: this function is not analytical, since it depends on the development of the merger-tree, from which we compute $L_X(\lambda_E)$ for any given $\lambda_E$. In particular, it has to be 
\begin{equation}
L_X (\lambda_E) = L_X^{\rm obs},
\end{equation}
where $L_X^{\rm obs}$ is the observed luminosity. Such equation is solved for the two $L_X^{\rm obs}$ values derived from observations of LBGs and QSOs by \cite{Cowie20} and \cite{Jiang09}, respectively.
}

The CDFs luminosity limits implies that MBH in LBGs must accrete at a low Eddington ratio\footnote{ As a test, we checked that a random scatter in the Eddington ration in the range $0-0.44$ introduces variations $<5$\% in the results.}, $\lambda_E = 3 \times 10^{-3}$; such value must increase to 0.36 to reproduce the quasar constraints. These values imply a somewhat different growth mechanism for MBH and SMBH. Accreted matter represents on average only 5\% of the final MBH mass at $z=6$, the rest being acquired by merging; however, its contribution raises up 
to 20\% for SMBH with mass $M_\bullet = 10^9 M_\odot$. In spite of such (mild) dependence on BH mass, a general conclusion is that accretion is a sub-dominant BH growth channel in S1 as a result of the large number of IMBH seeds available for mergers in this scenario. To gain further insight on this important aspect, we show in Fig. \ref{Fig02} with red lines the growth history of a MBH hosted by a LBG-type halo (left panel), and that for a SMBH (right). For this calculation, we followed the growth history of the seed located in the highest leaf from $z\approx 17$ down to the root.
{
In particular, for each halo our code keeps track of two masses:
(i) the mass of the merged seeds, $M_1$, (ii) the mass of directly accreted material, $M_2$. 
The actual BH mass is the sum of the two, $M_1+M_2$. At each time-step, we add the mass $M_2$, assuming that accretion occurs at a fraction $\lambda_E$ of the Eddington rate. When a merging event occurs, we sum the two masses, keeping separate track of the two quantities.
For example: (a) in a seeded leaf, $M_1 = 10^5 M\odot$ and $M_2 = 0$; (b) in the final root, $M_1$ is the sum of all the seeds in the merger tree. $M_2$ is the sum of mass accreted by each BH in the tree in all time-steps. In this case, the final mass of the black hole is $M_1 + M_2$.
}

For $f_{\rm seed}=1$ (S1) the growth is characterized by several vertical discontinuities associated with merging events, with accretion in between them playing a minor role, particularly at high redshift. Only $2.5\times 10^6 M_\odot$, out of the final MBH mass of $1.5\times 10^8 M_\odot$, have been accreted. For the SMBH the situation is only quantitatively different, with accretion along this branch contributing more ($20\%$) to the growth; however, mergings are still dominating the rise of the curve. 

\subsection{Inefficient seeding scenario}\label{S2}
In the S2 scenario $f_{\rm seed}=0.05$. This corresponds to a likely more physical situation in which only a minor fraction of candidate halos manage to form a IMBH seed. The best fit to the MBH -- halo mass relation (Fig. \ref{Fig01}, right panel) { is $\log \Mbh = -3.1^{+0.1}_{-0.1} + 0.95^{+0.01}_{-0.01} \log M_h$, } i.e., not too different from S1. The ratio $M_\bullet/M_h$ is also very similar to S1, apart from a slightly higher variance at low BH masses. { To be more quantitative, the BH mass in halos $10^{11-12} M_\odot$ is only 25 - 30\% larger in S2 with respect to S1.} These results are not particularly surprising as the two scenarios are bound to satisfy the same constraints.  
%
%
\begin{figure*}
\vspace{+0\baselineskip}
{
\includegraphics[width=1\textwidth]{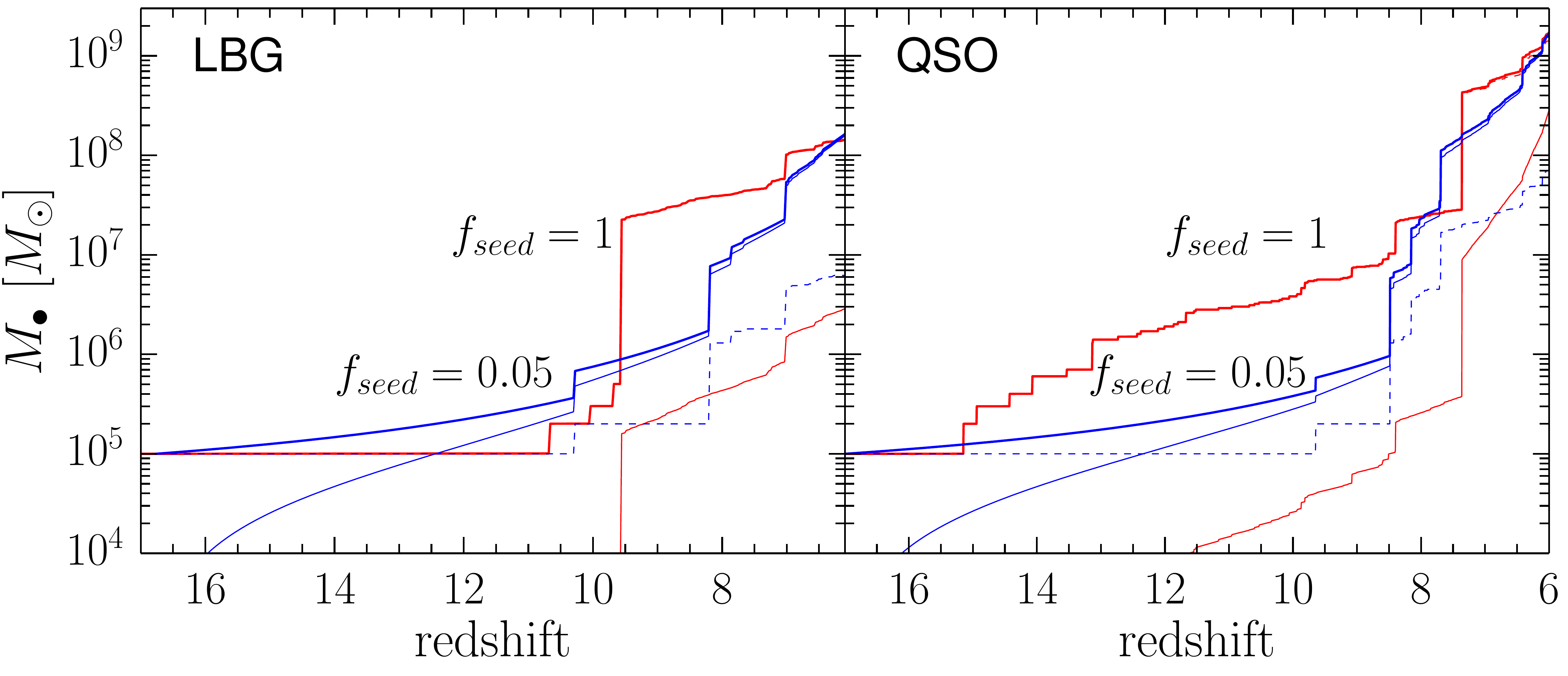}

}
\caption{Growth history of BH hosted by Lyman Break Galaxies (left) and quasars (right) as a function of redshift. In each panel we show the total BH mass (thick lines), and the mass contributed by accretion (thin) for the two scenarios S1 ($f_{\rm seed}=1$, red curves) and S2 ($f_{\rm seed}=0.05$, blue). For S2 we also show the contribution by mergers (dashed blue).}
\label{Fig02}
\end{figure*}

However, the key difference is that because of the scarcer availability of seeds, the required production of SMBHs requires a higher Eddington ratio, $\lambda_E = 0.22$ in order to accrete sufficient mass. However, if we force the LBG MBH (and lower mass BH) to accrete at the Eddington ratio required by the X-ray limits, $\lambda_E = 2\times 10^{-3}$, the early phase of the growth is strongly suppressed -- also lacking a major merger contribution, and would be too slow to climb up to the SMBH range. Hence, $\lambda_E$ has to be larger even in the MBH regime. For simplicity we have then assumed a constant $\lambda_E=0.22$ in the entire BH mass range.  Clearly, with this accretion rate LBG MBHs ({$\Mbh \simeq 10^{8.3} M_\odot$}) would be very luminous in X-rays. From eq. \ref{Lband} we obtain {$L_X =1.0\times 10^{44} \ergs$ (for $f_X=0.0153$}, averaged over the soft and hard X-ray bands, and for the appropriate bolometric luminosity), and hence largely exceeding the CDFs upper limit $L_X^{*} =3\times 10^{42} \ergs$. 

We are forced to assume that the X-ray flux from BHs in LBGs (and to a much smaller, but not negligible extent also in quasars) must be locally absorbed by intervening gas and dust. The transmitted fraction of the X-ray luminosity, $T_X$, can be determined\footnote{Strictly speaking, this is just a lower bound on the amount of absorption. In some cases, the central MBH might be obscured by even higher gas column densities, as found e.g. by \citet{Damato20} and \citet{Vito18} for 6 sources in the CDFs.} by imposing that $T_X L_X = L_X^*$, or $T_X=0.03$.  The required optical depth to $\approx 1$ keV photons to achieve such reduction is $\tau_X= - \ln T_X \simeq 3.53$, which for a solar metallicity gas implies an absorbing column  $N_H= 1.44\times 10^{22} \rm cm^{-2}$. We note that this conclusion perfectly agrees with LBG simulations at $z=6$, see e.g. Fig. 2 of \citet{Behrens19}. { This value is also consistent within 1$\sigma$ with those found by \cite{tre2019MNRAS.487..819T} and \cite{ni2020MNRAS.495.2135N}.}

The differences between S2 and S1 are also evident in the growth history of BHs hosted by LBGs and QSOs. Looking again at Fig. \ref{Fig02} we see that the $f_{\rm seed}=0.05$ blue curves are smoother, as a result of the more continuous growth associated with accretion\footnote{In reality a few jumps are seen also in the the accretions curves. These correspond to the nodes of two merger tree branches, where we sum the past accreted matter in each of the two.}. Indeed, both for LBG and QSO black holes, the growth is completely (97\%) dominated by accretion, with mergers playing a negligible role. This different balance between the two mechanisms entails an initial slower BH growth in S2. For example, the LBG  (left panel) BH at $z\simeq 8$ is about 20$\times$ less massive than predicted in S1, for which $\Mbh=4\times 10^7 \Msun$. The same effect is visible also in the QSO BH track (right panel), albeit shifted to a higher redshift range $z =10-12$. Eventually, the growth in S2 catches up with that of S1 by $z=6$. 

As the growth history encodes a memory of the initial seeding physics, it opens very interesting experimental perspectives to test when and where the first IMBH appeared on the cosmic stage. Proving the existence of $\Mbh > 10^7 \Msun$ BH in LBGs at $z\simeq 8$, for instance, would significantly favor an efficient seeding scenario, with relevant consequences on the production of UV photons in the early Universe.

\section{Massive black holes in LBGs}\label{sec:lum}
Having clarified the relation between MBHs and their host halo, to enable a meaningful observational comparison it is necessary to connect the properties of MBHs to those of the Lyman Break Galaxy population. To this aim we use available galaxy UV luminosity functions (LF) to associate a UV AB magnitude at $1375$\AA, $M_{\rm UV}$, to each halo. We then compute the MBH UV luminosity from our model, and finally combine the two in the total UV LF.
%
%
\begin{figure*}
\vspace{+0\baselineskip}
{
\includegraphics[width=1.\textwidth]{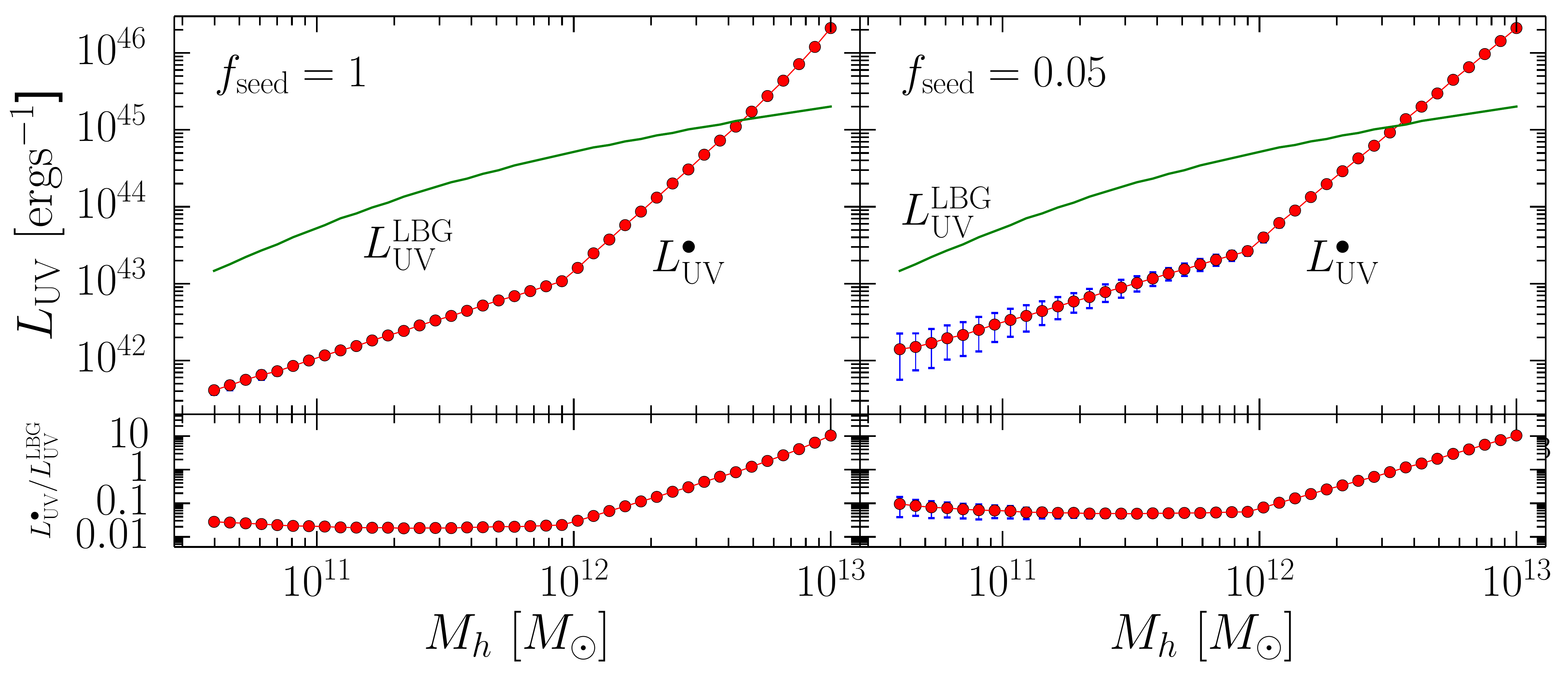}
}
\caption{\textit{Left panel}: LBG (green line) and BH (red points) UV luminosity vs. halo mass for S1. The bottom panel shows the luminosity ratio $L^\bullet_{\rm UV}/L^{\rm LBG}_{\rm UV}$ between the two components across the halo mass range. Variance of individual points is evaluated from $\approx 100$ merger tree realizations performed per halo mass bin. \textit{Right}: Same for S2 ($f_{\rm seed}=0.05$). 
}
\label{fig:luv}
\end{figure*}

\subsection{Galaxy UV luminosity}
\label{ssec:gallum}
\cite{bouwens2015} have studied $\approx 10400$  star forming galaxies at redshift $4 < z < 10$, and derived their UV LF. We then adopt their data at $z=6$, and perform an abundance matching analysis to associate a $M_{\rm UV}$ to each halo mass. This entails solving the following equation:
\begin{equation}
\int_{M_h}^{+\infty} \frac{dn}{dM_h'} dM_h' = \int^{M_{\rm UV}}_{-\infty} \frac{dn}{dM_{\rm UV}'}dM_{\rm UV}'.
\end{equation}
In the previous expression $dn/dM_h$ is the halo mass function \citep{shethtormen} implemented in the numerical code developed by \cite{mfpaper}; $dn/dM_{\rm UV}$ is the experimentally determined LF at $z=6$ \citep{bouwens2015}. 

The resulting relation between halo mass and the UV luminosity of the galaxy is reported in Fig. \ref{fig:luv}. For reference, the typical $M_h=10^{12} M_\odot$  LBG halo hosts a galaxy with UV luminosity of {$5 \times 10^{44}\ergs$}. Using the standard \citet{Kennicutt98} conversion factor of $4.46\times 10^{9} L_\odot/\Msun \rm yr^{-1}$, such luminosity corresponds to a star formation rate {${\rm SFR}=28.7\, \Msun \rm yr^{-1}$}.

{ For completeness, we derive also the X-ray luminosity produced by high-mass X-ray Binaries (hereafter, XRB), and associated with this SFR. Locally (Mineo et al. 2012; Mesinger 2015) the following relations holds:
\begin{equation}
    L_{\rm 0,XRB} = 3 \times 10^{39}  \left(\frac{\rm SFR}{M_\odot \, \rm yr^{-1}}\right) {\rm erg \, s^{-1}}.
\end{equation}
According to \cite{dij2012MNRAS.421..213D} and \citet{Lehmer_2016} the local relation evolves up to $z = 7$ as follows:
\begin{equation}
    L_{\rm XRB}(z) \approx L_{\rm 0,XRB}(1 + z);
\end{equation}
we refer to \cite{paper1} for more details. At $z=6$ this yields $L_{\rm XRB}= 6\times 10^{41} {\rm erg \, s^{-1}}$. Hence, the XRB luminosity is $\approx 0.2 L_X^*$. Note, in addition, that HMXB have a much softer spectrum with respect to AGN, and therefore most of their restframe luminosity is redshifted out of the Chandra bands.  For these reasons, we conclude that at best the HMXB impact on our results is small.}

\subsection{BH UV luminosity}\label{ssec:bhlum}
From the results obtained in Sec. \ref{sec:BH-halo} it is straightforward to compute the UV BH luminosity, using the derived $\Mbh$ and $\lambda_E$ values for the two scenarios, along with eq. \ref{Lband}, { and the bolometric correction by \cite{shen2020}}; this is displayed in Fig. \ref{fig:luv}. 
However, due to obscuration effects, the calculation for S2 requires an extra step. 

We have seen that in a typical LBG, X-ray emission must be absorbed by a gas column $N_H=1.44\times 10^{22} \rm cm^{-2}$. The corresponding optical depth at 1450\AA, adopting a Milky Way $R_V=3.1$ extinction curve \citep{Weingartner01} and solar metallicity, is $\tau_{\rm UV}=\sigma_{\rm UV} N_H = 4.8 \times 10^{-22} {\cal A}N_H = 19.4$, where ${\cal A}=2.75$ is the 1450\AA-to-V band attenuation ratio \citep{Ferrara19}. Differently from X-rays, whose opacity is dominated by gas photoelectric effects, (non-ionizing) UV photons mostly interact with dust by which they are absorbed and scattered. In spite of the large UV optical depth, scattering enables a varying fraction, $T_{\rm UV}$, of photons to escape from the system. $T_{\rm UV}$ depends on $\tau_{\rm UV}$, and on the optical properties of dust grains, namely the albedo, $\omega$, and the Henyey-Greenstein scattering phase function, $g$. The classical solution \citep{Code73} for a central source surrounded by a spherical gas/dust distribution obtained with the two-stream approximation, and confirmed by Monte Carlo radiative transfer simulations\footnote{Online digital data for the adopted configuration can be found at \url{www.arcetri.astro.it/~sbianchi/attenuation/E_MC.att}.} \citep{Ferrara99}, is appropriate here. This yields the transmitted UV fraction
\begin{equation}\label{TUV}
    T_{\rm UV} = \frac{2}{(1+\zeta)e^{\xi \tau_{\rm UV}} + (1-\zeta)e^{-\xi \tau_{\rm UV}} },
\end{equation}
where
\begin{equation}
\zeta     = \sqrt{(1-\omega)/(1-\omega g)} = 0.916, 
\end{equation}
and
\begin{equation}
\xi       = \sqrt{(1-\omega)(1-\omega g)} = 0.691, 
\end{equation}
having assumed the appropriate MW dust parameters $\omega=0.3668$ and $g=0.6719$ \citep{Weingartner01}. We then find $T_{\rm UV}=2\times 10^{-6}$ for a MBH hosted by a LBG halo. Such value is much larger than that for a pure absorption case $5.4\times 10^{-9}$, obtained by setting the albedo $\omega = 0$ in eq. \ref{TUV}. However, the gas is well known to be clumpy in the circumnuclear regions of AGN. \citet[][{ specifically see their Fig. 1}]{Bianchi00} showed that this situation leads to a lower effective $\tau_{\rm UV}$. They find that for a realistic case in which 75\% of the gas mass is in clumps (clumping factor $f_c=0.75$) the effective optical depth is $\tau_{\rm UV}^* \simeq\tau_{\rm UV}/3.5 = 5.55$. Then, the corresponding (effective) transmissivity is $T_{\rm UV}^* = 0.023$; note that, by chance, $T_{\rm UV}^* \approx T_X$. As a guide we will then assume this value when discussing S2 implications for LBGs in \ref{sec:impl}. 

Similarly, $T_{\rm UV}^*$ for QSO is derived by imposing that the bolometric luminosity in our most massive halo is $10^{47} \ergs$;  we linearly interpolate values of $T_{\rm UV}^*$ from the QSO to the LBG halo ranges, and keep it constant below the LBG halo mass. Finally, the emerging UV luminosity, as a function of the BH mass, is 
\begin{equation}
    L_{\rm UV}^\bullet = T_{\rm UV}^* f_{\rm UV} \lambda_E L_E.
\end{equation}

For the LBG, the predicted BH UV luminosity in S1 (S2) is $L_{\rm UV}^\bullet=1.5 \,  (3) \times  10^{43} \ergs$. The BH contribution to the total (galaxy + BH) luminosity increases with halo mass (the relative ratio of the two components is displayed in the bottom panel of the Fig. \ref{fig:luv}). 

In the LBG halo mass range $10^{11-12}\, \Msun$, the BH luminosity is $\simeq 1/50$ of the stellar one. However, this ratio raises in more massive halos until the BH outshines the host galaxy by a factor $\simeq 3-4$ in quasars. These results are qualitatively  the same for both scenarios. BHs in S2 are $\sim 5$ times more luminous compared to S1 in the LBG range.
This is mostly due to the different value of the product $T_{\rm UV}^* \lambda_E$.

For QSOs, S1 and S2  yield the same results, because they are both anchored to the QSO abundance constraints.
We recall that as in S2 $\lambda_E$ is about $100\times$ higher { in the LBG regime}, the CDFs X-ray and QSO constraints can only be satisfied if the BH emission in LBGs is heavily absorbed (see Sec. \ref{S2}).
The previous results suggest that UV luminosity of LBG is largely dominated by stars. Although we are not dealing with ionizing photons, this finding resonates with those by \citet{Trebitsch20} who conclude that faint AGN do not contribute significantly to cosmic reionization.  

\subsection{UV luminosity function}\label{UVLF}
\begin{figure*}
\vspace{+0\baselineskip}
{
\includegraphics[width=0.90\textwidth]{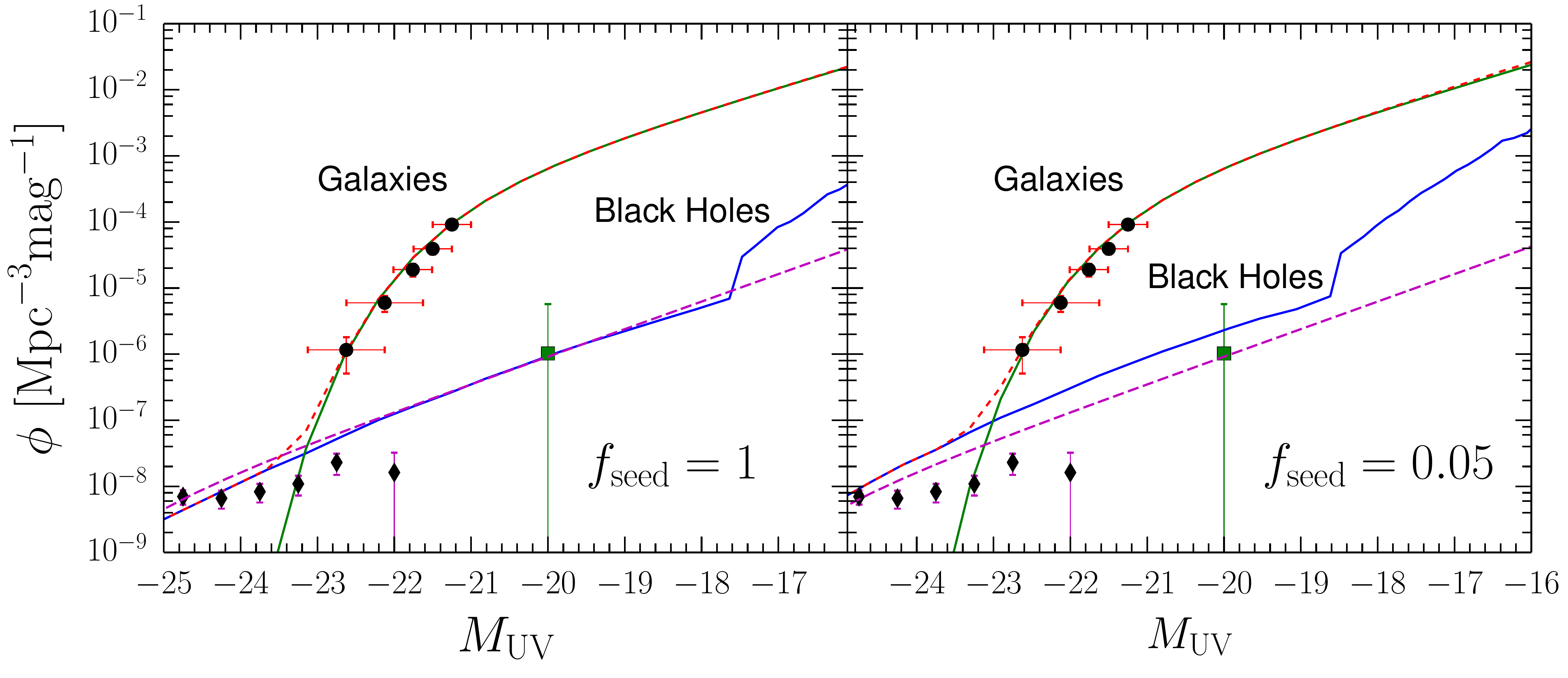}
}
\caption{\textit{Right}: Combined LBG and BH UV luminosity functions for scenario S1 ($f_{\rm seed}=1$) at $z=6$. The total (red dashed line) LF is the sum of the LBG (green) and BH (blue) contributions. Data points are from \citet{Bowler15} (circles), \citet{Matsuoka18} (diamonds) and \citet{parsa2018} (square). The long-dashed magenta line is the QSO LF best fit from \citet{Onoue17}.  \textit{Right}: Same for S2 ($f_{\rm seed}=0.05$). }
\label{fig:lumfun}
\end{figure*}

We are now ready to predict the BH contribution to the observed galaxy UV LF. This can be formally written as
\begin{equation}
  \phi \equiv \frac{dn}{dL_{\rm UV}^\bullet} = \frac{dn}{dM_h}\,\frac{dM_h}{d\Mbh}\, \frac{d\Mbh}{dL_{\rm UV}^\bullet},
\end{equation}
where ${dM_h}/{d\Mbh}$ and ${dM_{\rm BH}}/{dL_{\rm UV}^\bullet}$ are the BH-halo mass relation, and the BH mass dependence of the UV luminosity, respectively. Fig. \ref{fig:lumfun} shows the results of this calculation for S1 and S2, after a final conversion of the UV luminosity into an absolute AB magnitude, $M_{\rm UV}$, at 1375\AA.
%
%

In both scenarios the LF is largely dominated by stellar emission up to very bright magnitudes $M_{\rm UV} \simgt -23$. At luminosities fainter than this, the BH LF has a power law shape extending to $M_{\rm UV} \simeq -17.5$; at even fainter fluxes the BH LF becomes uncertain as the Eddington ratio is not constrained. The fraction of galaxies powered by a BH at $M_{\rm UV} \simeq -17.5$ is $10^{-3}$; this ratio increases to $6\times 10^{-3}$ (1.0) at $M_{\rm UV} \simeq -22$ ($-23.5$). The very bright end of the LF is dominated by rare ($\phi < 10^{-8} \rm Mpc^{-3}$) sources in which BH emission outshines star formation (i.e. quasars). There, the LF deviates from the Schechter function and becomes a power-law. 

These results agree well with available data from large survey such as the SHELLQ \citep{Matsuoka18}, GOLDRUSH \citep{Ono18}, and SHELA \citet{Stevans18}. SHELLQ, in particular, measured the quasar UV LF at $z \simeq 6$ over the wide mag range $-30 < M_{\rm UV} < -22$. The 
observed ratio of galaxies powered by a BH at $M_{\rm UV} \simeq -22 (-23.5)$ is $1.6\times 10^{-3} (1)$, in almost perfect agreement with our results (note that the measurement at the faintest $M_{\rm UV} < -22$ luminosity is affected by a considerable error). While the normalization of the LF at $M_{\rm UV}=-23.5$ also agrees with SHELLQ { (note that the BH and galaxy LFs overlap at this magnitude)}, predicting a BH density of $10^{-8} {\rm Mpc}^{-3}$, the faint-end slope\footnote{We follow \citet{Matsuoka18}, and define the slope from the power-law fit $\log(\phi/\phi^*) = {-0.4(\alpha+1)(M_{\rm UV}-M_{\rm UV}^*)}$.}, $\alpha$, of our { black hole} LF function is steeper. We find $\alpha = -2.5$, which must be compared with the SHELLQ best fit value $\alpha=-1.23^{+0.44}_{-0.34}$. We note that our faint-end slope is marginally consistent with that derived by \citet{Onoue17}, $\alpha=-2.04^{+0.33}_{-0.18}$, who included also X-ray detected AGN from \citet{Parsa18}, and the multi-redshift determination by \citet{Manti17}, $\alpha=-1.33^{+0.88}_{-0.93}$. This discrepancy likely indicates that a considerable fraction of AGN at $z=6$ is indeed obscured as we confirm here. Interestingly the two seeding scenarios cannot be disentangled purely from the LF. This is because they are both bound to satisfy the observational constrains at the LBG and QSO mass scales. However, we recall that -- in order to satisfy those constraints --  in S2 a large fraction of the accretion luminosity must be absorbed by gas and dust. This has important implications that we discuss in the next Section. 

\section{Implications and tests}\label{sec:impl}

{
The previous analysis suggests that the assumption that LBGs host MBH is consistent with observational constraints. 
}
In order to circumvent the tight limits imposed by X-ray observations, one has to assume that either such MBHs accrete at a very low rate (scenario S1), or their emission is obscured (S2). Although neither possibility can be discarded, we recall that S1 requires a, perhaps implausible, maximal efficiency of seed formation via the direct collapse mechanism. 
For this reason, we concentrate next on the implications of S2, and the possible ways to test them.

\subsection{Infrared emission}
We have seen that in a typical LBG, $\tau_{\rm UV}^*= 5.5$, implying that $>99.6$\% of the UV luminosity \textit{produced} by the MBH, $L_{\rm UV}^\bullet [(1-T_{\rm UV}^*)/T_{\rm UV}^*]  \simeq f_{\rm UV}\lambda_E L_{E}$, is absorbed by dust, and converted into thermal infrared emission. We recall from Sec. \ref{S2} that $\lambda_E=0.22$, $\Mbh=10^{8.3} \Msun$, and $L_E=3\times 10^{46} \ergs$; hence, the unobscured, intrinsic UV luminosity is $1.31\times 10^{45} \ergs \simeq L_{\rm FIR}$, where $L_{\rm FIR}$ is the total far infrared luminosity in the $8-1000\,\mu$m range.  To proceed further, we need to estimate the dust mass from the absorbing column\footnote{We have verified that the above predicted LBG luminosities are in perfect agreement with the observed $L_X - L_{\rm FIR}$ relation presented in Fig. 12 of \citet{Pouliasis20}.}
$N_H=1.44 \times 10^{22} \rm cm^{-2}$.   

We envisage two possibilities: (a) absorption is produced by a central obscurer local to the MBH, which we can tentatively identify with the dust torus, whose size we assume to be $R_H \simeq 1$ pc \citep{Netzer15}; (b) the absorbing dust is part of the interstellar medium of the host LBG { \citep{circo2019A&A...623A.172C}} . Numerical simulations \citep{Barai18} indicate that $N_H\simeq 10^{22} \rm cm^{-2}$ is found at a typical distance $R_H \approx 500$ pc from the center in AGN-host galaxies\footnote{Tentatively identified with the Narrow Line Region. Such $N_H$ corresponds to a mean gas density of $10\, \rm cm^{-3}$.} with a halo mass of $\approx 10^{12} \Msun$. The dust mass (assuming a dust-to-gas ratio ${\cal D}= 1/162$ \citep{Galliano08}) is then $M_d = (2.8, 7\times 10^5) \Msun$ for (a) and (b), respectively. 

The dust temperature, $T_d$ is the determined by the following expression \citep{Hirashita14}, which assumes a gray-body emission: 
\begin{equation}
    T_d = \left(\frac{f_{\rm UV}\lambda_E L_{E}}{\Theta M_d}\right)^{1/(4+\beta)};
\label{Td}
\end{equation}
where
\begin{equation}
    \Theta = \frac{8\pi}{c^2}\frac{\kappa_{158}}{\nu_{158}^\beta}\frac{k_B^{4+\beta}}{h_P^{3+\beta}}\zeta(4+\beta)\Gamma(4+\beta)=1.02\times 10^{-5},
\end{equation}
the mass absorption coefficient, $\kappa_\nu = \kappa_{158}(\nu/\nu_{158})^\beta$ is pivoted at a the reference wavelength of 158$\mu$m since high-$z$ ALMA observations are often tuned to the rest wavelength of [CII] emission. We take $\kappa_{158}=20.9\, {\rm cm}^2 {\rm g}^{-1}$, $\beta=2$ appropriate for graphite grains following \citet{Dayal10} and references therein; $\zeta$ and $\Gamma$ are the Zeta and Gamma functions, respectively; the other symbols have the usual meaning. We then obtain $T_d=(533,78)$ K for (a) and (b), respectively. As expected, dust located close to the MBH gets hotter. The peak wavelength for the gray-body adopted here is $\lambda_m = 0.29/T_d$, hence yielding $\lambda_m = (5.4, 43.1)\, \mu$m. For a LBG located at $z\simeq 6$, the redshifted emission peak nicely falls in the SW/SMI bands of SPICA for case (a); for case (b) the Rayleigh-Jeans portion of the spectrum is at reach of ALMA. It is then useful to compute the expected flux in these two cases. By applying the standard formula
\begin{equation}
    f_\nu = \frac{(1+z)}{d_L^2}\, \kappa_{(1+z)\nu} M_d B_{(1+z)\nu}(T_d),
    \label{fnu}
\end{equation}
we predict a flux of $(38, 27)\mu$Jy in the SPICA SW/SMI band and for ALMA Band 6; these fluxes are well at reach of these instruments. While SPICA is still in the planning phase, available ALMA continuum observations of $z\simeq 5-6$ LBGs at 158 $\mu$m indeed report fluxes that are comparable to the above one. For example, HZ6 a LBG at $z=5.3$ part of the \citet{Capak15} sample with $M_{\rm UV}=-22.5$, hence comparable to the reference LBG considered here, has a measured continuum flux of $129 \pm 36\, \mu$Jy. Hence, according to our results, $>16\%$ of the observed flux could be contributed by MBH accretion luminosity if the obscuring dust is located in the Narrow Line Region. 

In summary, hot dust is expected only if a MBH is present and the dust obscurer is local ($\simeq 1$ pc) to it; if absorption occurs on larger scales (comparable to the NLR, several hundreds pc) stars and MBH accretion contribute similarly to observed continuum flux. Hence, a MBH cannot be excluded by a SPICA non-detection; an ALMA detection cannot uniquely disentangle the MBH contribution from the stellar one. 


\subsection{UV emission lines}
To make progress, it is necessary to combine FIR probes with an unique feature of MBH accretion, such as UV emission lines. In particular, 
we should search for ionized species with an ionization potential $>4$ Ryd, which cannot be produced by even the hardest (e.g. binaries) stellar radiation sources, with the possible exception of elusive Pop III stars for \HeII\, \citep{Pallottini15}. The most suitable candidates are then the \NV\, 1240\AA\, (IP=77.47 eV) and \HeII\, 1640\AA\, (IP=54.4 eV) lines. \citet{Laporte17} recently reported a $\approx 5\sigma$ detection of these two lines in the redshift\footnote{We warn that our model is tuned to $z =6$, so uncertainties of about a factor of 2 (see Fig. \ref{Fig02}) in the BH mass and related quantities must be accommodated.} $z=7.15$ galaxy COSY ($M_{\rm UV}=-21.8$, SFR=$20.2\, \Msun \rm yr^{-1}$), opening the interesting possibility that this system might host a central MBH powering them. The measured rest-frame equivalent widths (EWs) is $3.2^{+0.8}_{-0.7}$\AA\, and $2.8^{+1.3}_{-0.9}$\AA, for \NV\, and \HeII, respectively. COSY is undetected in the 158$\mu$m continuum (upper limit $<14\, \mu$Jy), which, according to eq. \ref{fnu}, should imply a AGN-heated dust with temperature $T_d \simgt 90$ K, and distributed within 300 pc of the MBH.

\citet{Dietrich02} studied 744 Type 1 AGN in $0 < z < 5$, spanning nearly 6 orders of magnitude in continuum. They find that, almost independently of redshift, the EWs of most emission lines (including \HeII) significantly anti-correlate with the continuum strength (akin to the \quotes{Baldwin effect}); the \NV\, EW is instead almost independent of $L_{\rm UV}$. For our predicted \textit{observed} luminosity of $1.31\times 10^{45} T_{UV}^* = 3 \times 10^{43} \ergs$, their relations (Fig. 7 of their paper) indicate a EW of $30$~\AA\, and $20$~\AA\, for \NV\, and \HeII, respectively. Given that our MBH is significantly obscured, we need to correct these EWs for line absorption. 
A simple correction can be obtained by using the line escape probability, $\beta$, as a function of the medium optical depth (as both lines are in the UV, we assume the same optical depth, $\tau_{\rm UV}^*= 5.5$, derived in Sec. \ref{sec:lum}). Such formalism (see e.g. \citet{Netzer90} states that the probability for a line to escape from the system is 
\begin{equation}
    \beta \simeq \frac{1}{1+ 2\tau_{\rm UV}^*} = 0.083.
\end{equation}
After correcting for this effect the two EW become equal to $\simeq 2.49$\AA, for \NV\, and $\simeq 1.66$\AA, for \HeII\, in qualitative agreement (barred the many uncertainties) with the observed values, including their relative ratio. Although this simplistic treatment cannot represent a conclusive argument, it clearly points towards the possibility that indeed COSY hosts a MBH with properties similar to those predicted here.  

%
%
\begin{figure}
\vspace{+0\baselineskip}
{\includegraphics[width=0.45\textwidth]{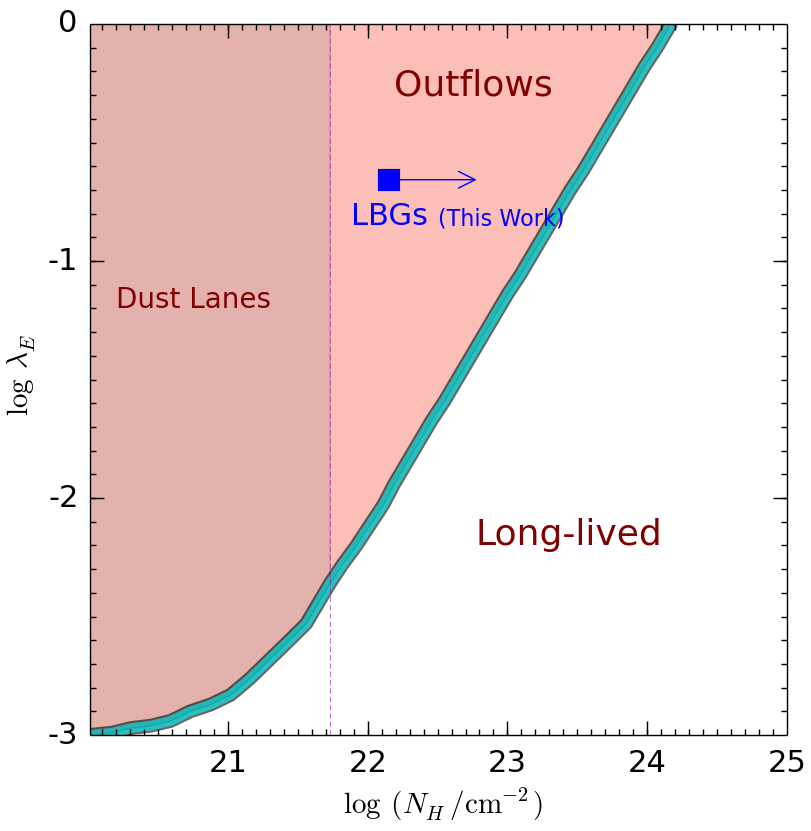}}
\caption{Regions in the $\lambda_E-N_H$ parameter space for a high-$\lambda_E$ AGN spectrum delimiting various regimes according to \citet{Fabian08} model. Systems lying to the left of the (maroon dashed) line develop powerful outflows if $N_H>21.7\, \rm cm^{-2}$; below that threshold filamentary structures (\quotes{dust lanes}) form. The predicted location of the typical, $M_{\rm UV}=-22$, LBG galaxy (blue point) at $z=6$ is also shown.}
\label{Fig05}
\end{figure}

\subsection{Outflows}
Accreting black holes might launch powerful outflows by converting their radiative energy into kinetic one. In a dusty medium, radiation pressure does not rely purely on Thomson scattering on electrons but it can additionally transfer momentum via an efficient coupling with dust grains. The amplification (or \quotes{boost}) factor of the radiation force in the UV bands is $A=\sigma_{UV}/\sigma_T \approx 1900$, where $\sigma_T$ is the Thomson cross-section, thus favoring the onset of radiation-pressure driven outflow from the galaxy. A proper treatment must include the frequency-weighting over the AGN spectrum. The calculation has been performed by \citet{Fabian08} as a function of the absorbing gas column density. In their model, they show that the boost factor is the inverse of the Eddington factor, $A=\lambda_E^{-1}$. 

Fig. \ref{Fig05} shows $\lambda_E\, {\rm vs.}\, N_H$ for a high-$\lambda_E$ AGN spectrum. Systems lying in the low $N_H$ -- high $\lambda_E$ region to the left of the curve develop powerful outflows, particularly if $N_H>21.7\, \rm cm^{-2}$, { this is consistent also with \cite{ni2020MNRAS.495.2135N}}. Lower columns  provide only a weak coupling to the radiation, leading only to the possible formation of filamentary structures (\quotes{dust lanes}). The predicted location of the typical, $M_{\rm UV}=-22$, LBG galaxy at $z=6$ (blue point in Fig. \ref{Fig05}) falls in the region in which outflows should develop. We recall that the $N_H$ derived from the CDFs X-ray data represents a lower limit, as indicated by the arrow. { If the ISM is clumpy, as assumed in Sec. \ref{ssec:bhlum}, the column density in the $x$-axis of Fig. \ref{Fig05} must be interpreted as referring to the clumps, which will therefore be individually accelerated. For a dedicated numerical work see, e.g. \citet[][]{Roth12}.}

Evidences for outflows in high-$z$ LBGs are rapidly accumulating, particularly thanks to the availability of ALMA observations. The original claim by \citet{Gallerani18} from a stacking analysis of the \citet{Capak15} sample, has been now confirmed by the ALPINE Large Program \citep{Ginolfi20, Fudamoto20}
Fast outflows have been tentatively identified in $z=5-6$ galaxies also by using deep Keck metal absorption line spectra \citep{Sugahara19}. 

Obviously, it might well be that these outflows are driven by supernova energy, rather than by an hidden AGN. \citet{Pizzati20}
showed that stellar outflows might explain the extended (size $\approx 10$ kpc) \CII halos observed around LBGs at $z=4-6$ \citep{Fujimoto19, Fujimoto20}. Interestingly, though, these authors noted that the required relatively large outflow loading factor, $\eta=3.2$, is only marginally consistent with starburst-driven outflows, and might instead indicate an additional energy input from a hidden AGN. Investigating the nature of outflows energy sources might lead to considerable progress in understanding the internal functioning of early galaxies, and their co-evolution with MBHs. 


\section{Summary}\label{sec:summary}
To address the possible presence of faint AGN powered by MBH in Lyman Break Galaxies in the Epoch of Reionization, we have run merger tree simulations implanted with direct collapse black hole seeds of mass $10^5 \Msun$ according to the prescriptions given in \citet{ferrara2014}. 
The BH growth, which can occur via BH-BH merging and matter accretion, is followed down to $z=6$ with an accretion rate determined by the Eddington ratio, $\lambda_E$, whose values is constrained by X-ray LBG and SMBH abundance/luminosity data. 
Depending on the seeded halo fraction, $f_{\rm seed}$, corresponding to different feedback-regulated formation efficiencies of direct collapse BHs, we consider (a) \textit{maximal seeding} ($f_{\rm seed}=1$, S1), and (b) \textit{inefficient seeding} ($f_{\rm seed}=0.05$, S2) scenarios.

The two scenarios predict a very similar $\Mbh/M_h \simeq 2\times 10^{-4}$ relation at $z=6$. This is not surprising as they are bound to satisfy the same observational constraints. However, they widely differ in many other properties. For example, in a typical LBG galaxy ($M_h =10^{12}\Msun, M_{\rm UV}=-22, n=10^{-5} \rm Mpc^{-3}$) accreted  matter  represents only 5\% of the final MBH mass, the rest being acquired by merging. Instead, in S2 accretion dominates in the (super-)massive BH range. It follows that, to satisfy X-ray constraints, the MBH luminosity in S2 must be obscured by an absorbing gas column density $N_H=1.44\times 10^{22} \rm cm^{-2}$, corresponding to a soft X-ray optical depth $\tau_X > 3.51$ (transmissivity $T_X=0.03$). For the QSO absorption is instead very small.  In addition, S2 predicts an initial slower BH growth: proving the existence of $\Mbh >10^7 \Msun$ MBH in $z\simeq 8$ LBGs, for instance, would significantly favor the maximal seeding scenario, with relevant consequences on the production of UV photons in the EoR.

We predict that the observed UV LF in both scenarios is largely dominated by stellar emission up to very bright mag, $M_{\rm UV} \simgt -23$, with BH emission playing a subdominant role. This finding is in agreement with the results by \citet{Volonteri17}. The fraction of galaxies powered by a BH at $M_{\rm UV} \simeq -17.5$ is $10^{-3}$; this ratio increases to $6\times 10^{-3}$ (1.0) at $M_{\rm UV} \simeq -22$ ($-23.5$). It is interesting to compare these predictions with available data for luminous LBGs. \citet{Capak15} carried out ALMA \CII\, observations of 10 LBGs at $z=5-6$; these sources have UV magnitudes in the range $-22.8 < M_{\rm UV} < - 21.5$. Among these, only 1 (HZ5) is classified as an AGN. We predict that the expected frequency of AGN in Capak's sample should have been $0.10 \pm 0.02$, in outstanding agreement with the 0.1 value found. Our results also are generally consistent with available QSO LF determinations, but the predicted faint-end slope $\alpha = -2.2$ is steeper than that derived by SHELLQ \citep{Matsuoka18}.

Although the two scenarios are both viable, S1 postulates a 100\% efficiency of seed formation.  As such,  for the more realistic S2 scenario in which MBHs grow by obscured accretion, we have explored the following implications:
\begin{itemize}
 \item[$\square$] \textit{Infrared emission} If the obscurer is local ($\simeq 1$ pc) to the MBH, the amount of dust implied is very small, $2.8 \Msun$; because of its high temperature, $T_d=533$ K, dust emission peaks at restframe $5.4\, \mu$m, and a typical $z=6$ LBG should produce a flux of  $38\, \mu$Jy in the SPICA SW/SMI band. If instead obscuration occurs on a scale typical of the NLR (500 pc), the larger mass ($7\times 10^5 \Msun$) of cooler ($T_d=67$ K) dust produces a $27\,\mu$Jy flux in ALMA Band 6. This represents $>16\%$ of the flux observed in similar LBGs, such as HZ6. 
 \item[$\square$] \textit{Emission lines} Although ALMA FIR continuum observations alone cannot conclusively pinpoint the presence of a faint AGN in LBGs, they can be combined with UV emission lines such as \NV\, and \HeII\, uniquely tracing AGN hard radiation. We show that the detected EWs of these two lines in COSY, a LBG galaxy at $z=7.15$ are successfully reproduced by our model, thus supporting the suggestion that COSY hosts  an obscured MBH with properties similar to those predicted here.
 \item[$\square$] \textit{Outflows} A MBH in a typical LBG galaxy, with the $\lambda_E$ and $N_H$ predicted here, should launch a powerful outflow according to the model by \citet{Fabian08}. This prediction is preliminarly supported by a number of recent findings, including the ALMA ALPINE survey, highlighting the presence of outflows that are only marginally consistent with starburst energetics, and might therefore require an additional energy contribution from a hidden, faint AGN.  
\end{itemize}

\section*{acknowledgements}
We thank F. di Mascia for useful discussions. AF acknowledges support from the ERC Advanced Grant INTERSTELLAR H2020/740120. Any dissemination of results must indicate that it reflects only the author’s view and that the Commission is not responsible for any use that may be made of the information it contains. Support from the Carl Friedrich von Siemens-Forschungspreis der Alexander von Humboldt-Stiftung Research Award is kindly acknowledged.

\bibliography{paper_bh}

\end{document}